\documentclass[12pt,preprint]{aastex}

\shorttitle{Negative ions in L1527}

\shortauthors{Harada \& Herbst}

\begin{document}




\title{Modeling Carbon Chain Anions in L1527}







\author{Nanase Harada}

\affil{Department of Physics, The Ohio State University, Columbus, OH 43210}

\author{Eric Herbst}

\affil{Departments of Physics, Astronomy, and Chemistry, The Ohio State University, Columbus, OH 43210}











\begin{abstract}
The low-mass protostellar region L1527 is unusual because it contains observable abundances of unsaturated carbon-chain molecules including C$_{\rm n}$H radicals, H$_{2}$C$_{\rm n}$ carbenes, cyanopolyynes, and the negative ions C$_{4}$H$^{-}$ and C$_{6}$H$^{-}$, all of which are more associated with cold cores than with  protostellar regions.  Sakai et al. suggested that these molecules are formed in L1527 from the chemical precursor methane, which evaporates from the grains during the heat-up of the region.  With the gas-phase  osu.03.2008 network extended to include negative ions of the families  C$_{\rm n}^{-}$, and C$_{\rm n}$H$^{-}$, as well as the newly detected C$_{3}$N$^{-}$, we modeled the chemistry that occurs following methane evaporation at $T \approx$ 25-30 K.  We are able to reproduce most of the observed molecular abundances in L1527 at a time of $\approx~5 \times10^{3}$ yr.   At later times, the overall abundance of anions become greater than that of electrons, which has an impact on many organic species and ions. The anion-to-neutral ratio in our calculation is in good agreement with observation for C$_{6}$H$^{-}$ but exceeds the observed ratio by more than three orders of magnitude for C$_{4}$H$^{-}$. In order to explain this difference, further investigation is needed on the rate coefficients for electron attachment and other reactions regarding anions.

\end{abstract}

























\section{Introduction}
L1527, an envelope of a low-mass star-forming region with IRAS04368+2557 at its center,  has the physical features of a class 0/I protostar, as discussed by \citet{andre00}. The infalling collapse of circumstellar gas was observed by \citet{ohashi97}, and the gas outflow was observed by \citet{tamura96}.  A number of interstellar molecules were discovered in L1527 as part of a molecular survey of dense cores by \citet{jorgensen04}.  Although one might expect the chemistry to be that of a hot corino \citep{cecc07}, the temperature of the envelope is $\approx$ 30 K, well below that of hot corinos (100 K).  More recently, \citet{sakai08a} detected unsaturated hydrocarbons and cyanopolyynes in the central core region of L1527. These species are unusual in corinos, and are more normally associated with cold ($T \approx 10$~K) dark clouds such as TMC-1 \citep{suzuki92,smith04}. 
\citet{sakai08a} explained the production of the carbon-chain species in terms of the evaporation of methane during the warm-up to 30 K followed by a rapid gas-phase synthetic chemistry, a process that they termed ``warm carbon-chain chemistry.'' This chemistry is related to a grain mantle desorption model for TMC-1 of \citet{markwick00}. In a recent paper, \citet{hassel08} used the osu gas-grain code to study the warm-up chemistry of L1527 in the vicinity of 30 K and found that the carbon-chain abundances could be reasonably reproduced in this ``lukewarm corino'' by a chemistry in which gas-phase methane is the precursor.  In addition to these unusual species, the two negative ion species C$_{4}$H$^{-}$ and C$_{6}$H$^{-}$ were also detected in L1527 \citep{sakai07,sakai08b,agundez08}.

   Anions in the interstellar medium were first discussed by \citet{dalgarno73}. \citet{herbst81} proposed that neutral radicals with large enough electron affinities can undergo efficient radiative attachment to form anions in cold dense clouds if they possess more than $\approx 4~$ atoms.  \citet{lepp88}  suggested that negatively charged polycyclic aromatic hydrocarbons (PAHs) could also form by radiative attachment.  \citet{wakelam08} modeled the gas-phase chemistry in dense cold cores with neutral and negatively charged PAHs, and found that the inclusion of PAH's allows good agreement with observations for small species with the use of so-called ``high-metal'' elemental abundances.  Confirmation of anions in interstellar and circumstellar sources awaited the measurement of their rotational spectra in the laboratory by Thaddeus and co-workers \citep{mccarthy06,gupta07}. In addition to the detections in L1527,  C$_{6}$H$^{-}$ and C$_{8}$H$^{-}$ \citep{mccarthy06,brunken07} were detected in TMC-1, while C$_{4}$H$^{-}$ \citep{cernicharo07}, C$_{6}$H$^{-}$ \citep{mccarthy06,kasai07}, and C$_{8}$H$^{-}$\citep{remijan07} were observed in IRC+10216. In response to these detections, the anions C$_{n}^{-}$, n=5-10, and C$_{n}$H$^{-}$, n=4-10, were added to the RATE06 (udfa.net) and IRC+10216 chemical networks \citep{millar00} by \citet{millar07} with estimated or calculated rate coefficients  \citep{terz00} for radiative attachment and photodetachment processes as well as measured rate coefficients for associative detachment reactions.  The augmented models were then used to study anionic abundances in TMC-1, IRC+10216, and photon-dominated regions such as the Horsehead Nebula.  \cite{millar07} found that, as predicted earlier by \citet{herbst81},  hydrocarbon radical anions can reach abundances exceeding 1\% of their neutral precursors, depending on their size. Another type of anion, C$_{3}$N$^{-}$, the radiative attachment of which was first studied by \citet{petrie97}, was also detected recently in IRC+10216 by \citet{thaddeus08}.  Its radiative attachment rate coefficient has been calculated anew by \citet{ho08}, with a result similar to that of \citet{petrie97} used here.
   
   In this paper, we re-study the gas-phase chemistry of L1527 with particular interest in the formation and depletion of anions and how their chemistry affects the overall  chemistry of the lukewarm corino.  The osu.03.2008 gas-phase network (see http://www.physics.ohio-state.edu/\verb+~+eric/) has been augmented to include formation and depletion reactions for the carbon cluster and  hydrocarbon radical anions as well as for C$_{3}$N$^{-}$.   The remainder of this paper is organized as follows.  In Section~\ref{code}, we discuss the gas-phase code used and the choice of initial abundances.  In Section~\ref{res}, the results are discussed and compared with observation for both anions and other species and with the earlier gas-grain results for non-anionic species.  In Section~\ref{var}, we consider how our results respond to variations in initial abundances and to the exclusion of anionic chemistry.  Finally, we end with a summary.

\section{Gas-Phase Code and Initial Conditions}
\label{code}
 The recent osu gas-phase network (http://www.physics.ohio-state.edu/\verb+~+eric/)  osu.03.2008 has been augmented by the addition of several families of negative ions and classes of reactions to form and destroy them.  The new negative ions included in the model consist of the carbon chain families C$_{\rm n}^-$ 
 (n = 5-10) and C$_{\rm n}$H$^{-}$  (n = 4-10) and the newly detected anion C$_{3}$N$^{-}$.  The reaction classes include radiative attachment (A + e $\rightarrow$ A$^{-}$ + h$\nu$), photodetachment (A$^{-}$ +  h$\nu \rightarrow$ A + e), dissociative attachment (AB + e $\rightarrow$ A$^{-}$ + B), associative detachment (A$^{-}$ + B $\rightarrow$ AB + e), anion-neutral reactions (A$^{-}$ + B $\rightarrow$ C$^{-}$ + D) and mutual neutralization (A$^{-}$ + B$^{+}$ $\rightarrow$ A + B).  Most of  these classes of reactions are contained in the  modified RATE06 network used in \citet{millar07}.  We have updated the rate coefficients for the radiative attachment of C$_{4}$H and C$_{5}$H according to the theory of 
 \citet{ho08}.  A large number of mutual neutralization reactions involving all anions and 57 major positive ions were added with rate coefficients $k = 7.5 \times 10^{-8}(T/300)^{-0.5}$ cm$^{3}$ s$^{-1}$ \citep{smith78,wakelam08}; without these reactions,  the correct balance of anions, electrons and positive ions cannot be maintained  for L1527 at later times, when the overall anionic abundance exceeds the electron abundance.    Newly added reactions for the formation and depletion of C$_{3}$N$^{-}$ and their rate coefficients are listed in Table \ref{cccn}.   Overall, the augmented network for L1527 contains 5941 reactions involving 469 species, an extension of osu.03.2008, which contains  4479 reactions involving 468 species.


We follow the gas-phase chemistry after its  warm-up from a cold cloud with icy mantles to a temperature in the range $25 - 30$ K,  by which most of the methane ice in grain mantles has evaporated along with other volatile species such as CO and N$_{2}$, which evaporate at slightly lower temperatures than methane.  We use a temperature of 30 K, a total density $n_{\rm H} = 10^{6}$ cm$^{-3}$ and visual extinction \(A_{\rm v}\)=10 mag \citep{sakai08a,jorgensen04}.  To convert observed column densities into fractional abundances, we use a column density $N_{\rm H} = 6 \times 10^{22}$ cm$^{-2}$ \citep{jorgensen04}.  The cosmic ray ionization rate $\zeta$ is taken to be $1.3 \times 10^{-17}$ s$^{-1}$. The initial abundances, shown in Table~\ref{abund}, are based on the gas-grain code results by \citet{hassel08}, in which the gas and grain populations during the warm up are calculated.    Since most of the water ice resides on grain mantles until much higher temperatures, the elemental carbon-to-oxygen abundance ratio in the gas is carbon rich, as can be seen by adding up the carbon-containing and oxygen-containing species.  The important precursor methane has a fractional abundance of $3 \times 10^{-6}$, in reasonable agreement with values obtained for cold ices using $ISO$ \citep{gibb04} and $Spitzer$ \citep{oberg08}.  The low abundances for metals reflect the fact that most are still embedded in grain mantles.
The chemistry starts quickly, as methane is partially converted into acetylene and more complex hydrocarbons via ion-molecule reactions \citep{hassel08}.


\section{Results}
\label{res}
\subsection{Hydrocarbon Radicals and  Anions}
Hydrocarbon radicals of the general linear structure C$_{\rm n}$H are the main precursors to the C$_{\rm n}$H$^{-}$ anions through radiative attachment. The calculated fractional abundances (with respect to $n_{\rm H}$) of C$_{4}$H, C$_{6}$H, C$_{8}$H, and C$_{10}$H along with their respective anions as functions of time are shown in Figure~\ref{fig:cnh-}.   For the first $3 \times 10^{4}$ yr, the neutral radicals increase in abundance with abundances inversely proportional to size.  At 10$^{3}$ yr, the main production channels of C$_{\rm n}$H are dissociative recombination from  precursor ions C$_{\rm n}$H$_{\rm m}^{+}$:
\begin{equation}
{\rm  C_{n}H_{m}^{+} + e^{-} \longrightarrow C_{n}H + ...},
\end{equation}
and neutral-neutral reactions from  smaller hydrocarbons:
\begin{equation}
\label{carb}
{\rm C +  C_{n-1}H_{2}   \longrightarrow C_{n}H + H.}
\end{equation}
The main formation reactions after 10$^{4}$ yr become the associative detachment processes 
\begin{equation}
\label{bare_anions}
{\rm C_{n}^{-} + H \longrightarrow C_{n}H + e^{-}}
\end{equation}
because the concentrations of the bare cluster anions become significant.  When peak values are achieved, at a time of $\approx 3-5 \times 10^{5}$ yr, the size vs fractional abundance order has broken down, and the most abundant radical becomes C$_{8}$H. The radical C$_{6}$H is next, while the still lower abundances of C$_{4}$H and C$_{10}$H are similar.  As steady state is achieved at still later times, the abundances of the three smallest radicals are nearly the same, while C$_{10}$H is somewhat lower.  The unusual dependence on size is due partially to carbon-rich elemental abundances and partially to the inclusion of anions, which increase the synthetic power of the chemistry and cause an ``edge'' effect, as discussed below in Section~\ref{anioneffect}.

The calculated fractional abundances of the C$_{\rm n}$H anions, depicted in Figure~\ref{fig:cnh-},  show an unusual dependence on size as well.  At times after $4 \times 10^{4}$ yr, the most abundant anion is C$_{8}$H$^{-}$, which slightly exceeds the smaller C$_{6}$H$^{-}$.  There is a gap in abundance between these two, and the pair of ions 
C$_{10}$H$^{-}$ and C$_{4}$H$^{-}$, which trail by a factor of 3-5.  The relatively low concentration of C$_{4}$H$^{-}$ is caused by the relative slowness of the radiative attachment reaction \citep{ho08}
\begin{equation}
{\rm  C_{4}H + e^{-} \longrightarrow C_{4}H^{-} + h\nu },
\end{equation}
despite the fact that the calculated value of the rate coefficient is known to be too large \citep{ho08}.
The observed fractional abundances of C$_{4}$H$^{-}$ and C$_{6}$H$^{-}$ in L1527 are $1.8 \times 10^{-13}$ and $9.7 \times 10^{-13}$ respectively \citep{sakai07,sakai08a}.  It can be seen that while the calculated value for the larger anion is close to the observed value at relatively early times ($\approx 10^{3-4}$ yr), the calculated abundance for the smaller anion is far too large at all but the shortest times.

\subsection{Cyanopolyynes and C$_{3}$N$^{-}$}
 Cyanopolyynes are produced via a variety of ion-molecule and neutral-neutral channels.  The existence of the C$_{\rm n}$H$^{-}$ anions boosts the efficiency of one production mechanism:
 \begin{equation}
 \label{cyano}
{\rm CN  +  C_{2n}H_{2} \longrightarrow HC_{2n+1}N + H}
\end{equation}
by enhancing the abundance of the the precursor C$_{\rm 2n}$H$_{2}$ through the associative detachment reactions
\begin{equation}
\label{c2nh2}
{\rm C_{2n}H^{-}  +  H  \longrightarrow  C_{2n}H_{2}  +  e^{-}.}
\end{equation}
Although cyanoacetylene, HCCCN, can be formed by neutral-neutral channels, its higher energy isomers, such as HNCCC, are thought to be formed only through dissociative recombination \citep{osamura99}; viz.,
\begin{equation}
{\rm C_{3}H_{2}N^{+}  +  e^{-} \longrightarrow HNCCC + H.}
\end{equation}
This particular isomer is important, because as can be seen in Table~\ref{cccn}, it leads to the anion C$_{3}$N$^{-}$ via an exothermic dissociative attachment reaction with hydrogen atoms.  This synthesis can be competitive with the more simple radiative attachment depending on the source.   Indeed, in L1527, it is the dominant pathway, as indicated in Figure~\ref{fig:C3N-}, which shows the calculated fractional abundances of HC$_{3}$N, HNCCC, C$_{3}$N and C$_{3}$N$^{-}$ as functions of time.  It can be seen that the abundance of the radical C$_{3}$N exceeds that of the isomer HNCCC for all times past 2$\times$10$^{3}$ yr.  However, the abundance of the radical does not exceed that of the isomer by more than a factor of 3, while the radiative attachment rate coefficient for C$_{3}$N$^{-}$ + e$^{-}$ is lower than the dissociative attachment coefficient for HNCCC + e$^{-}$ by a  factor of 100, as shown in Table~\ref{cccn}.  The predicted fractional abundance of the anion lies somewhere in the range 10$^{-11} - 10^{-12}$ for most times.  Although this range of values seems very low, it should be noted that the anion C$_{4}$H$^{-}$ was detected with an even lower fractional abundance.

\subsection{Anion-to-Neutral Ratio}
The calculated and observed anion-to-neutral abundance ratios vs time are shown in Figure~\ref{fig:anion_ratio} for C$_{4}$H and C$_{6}$H. Despite the fact that both anion and neutral are significantly overproduced at times $> 10^{4}$ yr, the calculated C$_{6}$H$^{-}$/C$_{6}$H ratio always lies close to the observed value of $\approx$ 10\%, exceeding it  by at most a factor of five at early times, and dipping below by less than a factor of two before steady state is reached.  This level of agreement suggests that the theoretical rate coefficient for the radiative attachment of C$_{6}$H \citep{ho08} is reasonably accurate although the alternative possible formation route via dissociative attachment of the carbene H$_{2}$C$_{6}$ is not included in our network \citep{sakai07,ho08}.  The  ratio C$_{4}$H$^{-}$/C$_{4}$H, as for other sources, is calculated to be much too large, due principally to the failure of the phase space treatment for  the radiative attachment of C$_{4}$H \citep{ho08}.


In Figure~\ref{fig:cmp_sum}, we plot the fractional abundances of the electron, total positive ions, and total anions.  After $4 \times 10^{4}$ yr, the total anion abundance exceeds the electron abundance, and anions become the dominant form of negatively charged particles. Since the mutual neutralizations of positive ions and negative ions are less rapid than dissociative recombination reactions between positive molecular  ions and electrons, negative ions are likely to stay more abundant than electrons.  Indeed, at steady state, the anionic abundance exceeds the electronic abundance by a factor of more than two.   Since anions can be more abundant than electrons, it was imperative to add a large number of mutual neutralization reactions to make sure that the correct proportion of anion, electron, and positive ions was obtained.

\subsection{Overall Comparison with Observation}
To determine the best fit to the observed 23 abundances for the anions and other species in L1527, we used the mean confidence level method of \citet{garrod}.  In this method, a confidence level 
$\kappa_{i}$ is defined for each species $i$ by the equation
\begin{equation}
\kappa_{i}=erfc \left( { | \log (X_{i}) - \log (X_{\rm{obs},i}) | } \over  { {\sqrt{2}}\sigma } \right),
\end{equation} 
where $X_{i}$ is the computed fractional abundance, $X_{\rm{obs},i}$ the observed fractional abundance, and $\sigma=1$.  The agreement is perfect for $\kappa_{i} = 1$, a factor of three departure from observation if $\kappa_{i} = 0.63$, and a one order-of-magnitude departure from observation if $\kappa_{i} = 0.32$.   We then take $\kappa$, the average value of $\kappa_{i}$, as the criterion for agreement.   In our network of reactions, we do not make a distinction between the acetylenic forms HC$_{\rm n}$H and the carbenes H$_{2}$C$_{\rm n}$.  Since only the latter are detected, we make the assumption that they represent only 1\% of the total C$_{\rm n}$H$_{2}$ abundance \citep{park06}.
The optimal value of $\kappa$ is 0.652 corresponding to an average agreement within a factor of three; this agreement occurs at a time of $4.8 \times 10^{3}$ yr.  The species and their observed and optimal calculated fractional abundance are shown in Table~\ref{obs.v.calc}, where it can be seen that 18 of the 23 molecules have computed abundances within an order of magnitude of the observed values, while 15 have computed abundances within a factor of three of the observed values.  The most extreme outlier is the anion C$_{4}$H$^{-}$, which is predicted to have an abundance higher  than observed by a factor of 61.  The other molecules that disagree with observation by more than an order of magnitude are C$_{4}$H, N$_{2}$H$^{+}$, and CN. Another anion, C$_{6}$H$^{-}$ is overproduced by an order of magnitude, C$_{5}$H and  HNC are  overproduced by more than a factor of a few,  while HCO$_{2}^{+}$ is underproduced by factor of five. The mean confidence limit is not a strong function of time despite the fact that the abundances of carbon-chain molecules tend to keep increasing past the optimal time; at steady-state, which occurs at times slightly in excess of 10$^{6}$ yr, the value of $\kappa$ is still 0.474, corresponding to an average discrepancy factor of slightly more than five.  At  steady-state, 13 species are in agreement within an order of magnitude, and 9 of them are in agreement within a factor of 3. The main difference between the optimal time and steady-state results concerns the abundances of carbon chain species, which become much larger than their observational values past the optimal time. In fact, at 5$\times$10$^{5}$ yr, when these molecules achieve their peak abundances, the confidence level is 0.411; a little worse than at steady-state.

If we exclude the anions from our network, and fit for the remaining 21 molecular abundances in L1527, we find an optimal value for $\kappa$ of 0.628 at $5.9 \times 10^{3}$ yr, corresponding to an average factor of three discrepancy, and a steady-state value of 0.487, corresponding to an average factor of five discrepancy.  The optimal results without anions are also tabulated in Table~\ref{obs.v.calc} as are the optimal results of the gas-grain warm-up model to 30 K \citep{hassel08}.  This latter model, which does not contain anions,  achieves an optimal agreement of a factor of three discrepancy when the temperature has risen to 25 K from its initial value of 10 K.  The time spent up to 25 K when the  gas-phase carbon chemistry can occur is roughly $1-2 \times 10^{4}$ yr.     Comparison of the gas-phase models with the gas-grain model shows that the gas-phase model without anions  achieves somewhat closer agreement with the gas-grain model, because, as discussed later, the anions do affect the chemistry.  It should also be noted that the observed molecular abundances towards L1527 excluding anions can also be fit to within an order of magnitude by a cold model, especially the non-carbon-chain species, which may not be spatially correlated with the carbon-chain molecules \citep{hassel08}.  

Predicted abundances for a wide class of molecules using the optimal times of the gas-phase fit with  anions as well as the gas-grain warm-up model to 30 K are listed in Table~\ref{predict}.  Most optimum gas-phase abundances tend to be only slightly smaller than their gas-grain counterparts.  Other than the negative ions, some significant disagreements ($>$ 1 order-of-magnitude)  concern the hydrocarbon radicals, which are depleted by electron attachment in the gas-phase model, the carbenes, and species such as methanol, which are not produced efficiently in the gas phase.  Despite the fact that the gas-grain prediction for methanol is much higher than the gas-phase result, methanol is still predicted to be of very low abundance in the gas until the temperature warms up to hot core levels \citep{hassel08}.

\section{Response to Variations}
\label{var}

\subsection{Varying Initial Conditions - Oxygen and Methane}
Unlike the gas-grain approach of \citet{hassel08} to L1527, where the evaporation of the volatile mantle species is determined by the chemical and physical processes, in the gas-phase approach we can vary the initial abundances from those listed in Table~\ref{abund}, which derive from evaporation from grain mantles in the warm-up gas-grain model.  It is useful to be able to vary the initial abundances because the warm-up scenario is approximate at best.  Here we consider variations in the initial abundances of atomic  oxygen  and methane (CH$_{4}$).  The C/O elemental abundance ratio is 1.06 with our standard abundances; an increase in O will move the ratio closer to unity.  The amount of methane is important to the warm carbon chemistry, since it is the precursor for all of it, and since its observed prior abundance in ices can vary \citep{oberg08}.

Table~\ref{ovary} shows the effect of changing the initial abundance of atomic oxygen on the time of best agreement,  the average confidence level $\kappa$, and the number of species within order-of-magnitude agreement with observation.  Although the O abundance is not large,  O atoms tend to destroy organic ions and radicals.  Varying the fractional O abundance from $5 \times 10^{-8}$ to $1 \times 10^{-6}$ shows that the optimal value for $\kappa$ peaks for $1-2 \times 10^{-7}$, although the criterion of number of molecules fitted to within an order-of-magnitude peaks at $5 \times 10^{-7}$. The sensitivity to changes in O abundance, however, is minimal.  On the other hand, the time of best agreement increases with increasing initial abundance of O.   This latter dependence arises from the fact that O slows down the synthetic process.


Regarding methane, the results for time of best agreement, $\kappa$, and order-of-magnitude agreement are shown in Table~\ref{methvary}.  There is little dependence on the optimal fit defined by either the $\kappa$ criterion or the number-of-molecules criterion as the initial methane abundance is raised from $3 \times 10^{-7}$ to $3 \times 10^{-5}$.  Here, though, the time of optimal agreement decreases with increasing methane abundance.
 Presumably the higher abundance of methane leads to the more rapid production of carbon-chain species. 
 



\subsection{The Effect of Anions}
\label{anioneffect}

It is interesting to compare the chemistry of our gas-phase model for L1527 with and without anions in more detail. Given that the overall abundance of anions in the model exceeds that of electrons at times later than $4 \times 10^{4}$ yr, one would expect the effect of anions to be more pronounced at such times.  By and large, this inference is true although for some classes of molecules there is little effect at all at earlier times, while for other species the effect is already a large one.  In Figure \ref{fig:carbon_chain_l1527}, we plot the fractional abundances of four classes of molecules  vs time -- the C$_{\rm 2n}$H radicals (Upper left panel),  the hydrocarbons C$_{\rm n}$H$_{2}$, consisting mainly of polyynes (HC$_{\rm 2n}$H) with small abundances of carbenes H$_{2}$C$_{\rm 2n}$ (Upper right panel), the bare carbon clusters (Lower left panel), and the cyanopolyynes HC$_{\rm 2n+1}$N ( Lower right panel)  -- with and without anionic chemistry.  Let us start with the hydrocarbons.   Here it can be seen that the effect caused by the anions increases with time, as expected, so that by steady-state the abundances computed without anions are mainly significantly lower than those computed with anions.  The reason is that the C$_{\rm n}$H anions increase the abundances of the hydrocarbons via reaction~(\ref{c2nh2}) as well as the analogous processes for odd numbers of carbon atoms.
We also note that at later times, including steady state, the largest hydrocarbon depicted, C$_{8}$H$_{2}$, has the largest abundance in the model with anions, and the smallest abundance in the model without anions.   The chemistry leading to this unusual result is complex and will be the subject of a future paper.  For now, we note only that the existence of anions has a catalytic effect on the production of more complex hydrocarbons from simpler ones.  For example, starting with the hydrocarbon C$_{6}$H$_{2}$, reaction with carbon atoms leads to the species C$_{7}$H (see reaction~(\ref{carb})), which is efficiently converted to its anion.  The anion, in turn, undergoes an associative detachment reaction with atomic hydrogen to form the more complex hydrocarbon C$_{7}$H$_{2}$ (see reaction~({\ref{bare_anions})).  The process continues through the largest hydrocarbon in the model.
 The enhanced abundance of this largest hydrocarbon suggests that still larger species must be included in the model to avoid an ``edge'' effect for potentially observable species, in which the concentrations of the larger species are over-estimated due to a paucity of destruction reactions.

If we look at the cyanopolyynes, the result, though more complex, is similar to that of the hydrocarbons.  The enhanced synthetic power with anions is doubtless caused by the enhanced production of these hydrocarbons, which can react with the radical CN to produce cyanopolyynes, as shown in reaction~(\ref{cyano}).  
The late-time abundance of the largest cyanopolyyne shown, HC$_{13}$N, is comparable with those of smaller cyanopolyynes.  
On the other hand, the plot of the C$_{\rm 2n}$H radicals shows that the existence of anions decreases their abundances starting at rather early times.   This effect is surprising because bare carbon anions are eventually involved in their formation (see reaction~(\ref{bare_anions})), but presumably destruction via the  formation of anions is more important.  For the bare carbon clusters, destruction by the formation of anions causes a similar effect.

\section{Summary} 

L1527 is an unusual source because the unsaturated carbon-chain molecules detected there are more normally associated with cold dense cores at lower temperatures.  \citet{sakai08a} proposed that the carbon-chain chemistry starts soon after rising temperatures allow the volatile species methane to evaporate into the gas.  Our group has studied this chemistry via two types of models.  In a previous model, \citet{hassel08} used the osu gas-grain code to study the chemistry in a cold dense core followed by warm-up to temperatures associated with L1527 ($T \approx 30$ K) and found a rich chemistry leading to carbon-chain species, as hypothesized by \citet{sakai08a}.  Comparison with observation showed that although the non-carbon-chain species in L1527 could be fit well either as remnants of a previous cold phase or current denizens of the warm-up phase, the carbon-chain species are fit significantly better with the warm model.  Although the gas-grain approach was quite useful, this method does not contain any anionic chemistry, and the chemistry is fixed by what happens in the earlier cold epoch and the rising temperatures.  We have followed this prior study with the isothermal ($T = 30$~K)   gas-phase study reported here because it allows us to include the gas-phase anionic chemistry without worrying about what happens when anions collide with grains and because it allows us to vary some of the initial abundances, which are mainly based on the earlier model.  The abundance of methane, for example, has been studied in the ice phase of cold sources \citep{oberg08} prior to evaporation and  found to depend somewhat on the source investigated. 

Including the chemistry of carbon-chain anions in our gas-phase network  has led to a large increase in the number of reactions, mainly because of the many cation-anion neutralization reactions needed.    With the chosen initial abundances, the best agreement with all 23 observed molecules in L1527 occurs at a time of  $4.8 \times 10^{3}$ yr, when the abundances of most carbon-chain species are increasing.   This agreement corresponds to an average factor of three discrepancy between observational and calculated values.  Variation in the initial abundances of either atomic oxgyen or methane changes the time of best agreement mainly because these species regulate both the formation and depletion rates of carbon-chain species.  For the observed anions, C$_{4}$H$^{-}$ and C$_{6}$H$^{-}$, we fit both the fractional abundance and the anion-to-neutral ratio reasonably well for the larger anion, but dramatically overproduce these quantities for the smaller anion.  Finally, it is to be noted that the steady-state abundances in the gas-phase model are reached by 10$^{6}$ yr, so that the chemical processes of interest in this paper are completed well before the gas-grain models show declines in the gas-phase abundances \citep{hassel08}.

A comparison of the chemistry occurring with and without negative ions shows that anions have a major effect on the chemistry of other species, especially at later times than that for the optimal fit, when the overall anionic abundance exceeds that of electrons.  This effect can either lower abundances of a given class of molecules or raise them, depending upon whether the formation of anions tends to deplete other species or enhance their production.  There is also a strange effect in which the dependence of the abundance of a molecule in a family such as the cyanopolyynes on size is changed when anions are included in the chemistry.  Typically, in the absence of anions, an increase in size results in a drop in peak abundance.  The inclusion of anions appears to make this dependence significantly smaller and even to vitiate it completely in some instances.  As will be shown in subsequent work, this effect is also  pronounced in the cold core TMC-1 under oxygen-rich conditions.  With relatively large abundances for the largest carbon-containing species in our network, it may well be necessary to add still larger molecules to the model, either by including all of the formation and depletion reactions that govern their chemistry, as in the IRC+10216 network \citep{millar00}, or by some more approximate approach \citep{herbst91}.  In addition, despite the size of the current network, the role of negative ion-neutral reactions remains to be fully explored \citep{cordiner08,cordmill08}.


\acknowledgments
We would like to thank Nami Sakai and Satoshi Yamamoto for initiating the problem, and for useful discussions.  We also thank George Hassel for his help in explaining the results of the gas-grain code.  E. H. acknowledges the support of the National Science Foundation for his research program in astrochemistry and the support of NASA for studies in support of space telescopes and the study of preplanetary matter.

\clearpage

 \begin{table}
 \caption{Formation and Depletion Reactions for C$_{3}$N$^{-}$}
 \label{cccn}
 \begin{tabular}{lrr}
 \hline\hline
 Reaction &  Rate Coefficient (cm$^{3}$ s$^{-1}$) & Reference \\
 \hline
 C$_{3}$N  +  e  $ \rightarrow$ C$_{3}$N$^{-}$  +   h$\nu$ & $2.0 \times 10^{-10}(T/300)^{-0.5}$ & 1 \\
 HNCCC + e $\rightarrow$ C$_{3}$N $^{-}$  +  H     &  $2.0 \times 10^{-8}(T/300)^{-0.5}$ & 2\\
 C$_{3}$N$^{-}$ + H $\rightarrow$ HC$_{3}$N + e & $1.0 \times 10^{-9}$ & 3\\
  C$_{3}$N$^{-}$ + C $\rightarrow$ C$_{4}$N + e & $1.0 \times 10^{-9}$ & 3\\
    C$_{3}$N$^{-}$ + O $\rightarrow$ C$_{2}$N + CO + e & $1.0 \times 10^{-9}$ & 3\\
    C$_{3}$N$^{-}$ + h$\nu$ $\rightarrow$ C$_{3}$N + e & $2.0 \times 10^{-9} \exp{(-2.0A_{\rm v})}$ & 4 \\
    C$_{3}$N$^{-}$ + C$^{+}$ $\rightarrow$  C$_{3}$N + C & $7.5 \times 10^{-8}(T/300)^{-0.5}$ & 5\\
\hline
\end{tabular}
\tablerefs{
(1) \citet{petrie97}\\
(2) estimate based on \citet{adams86} \\
(3) based on \citet{eichel07}\\
(4) based on \citet{millar07} \\
(5) sample positive ion; see \citet{smith78} for typical rates
}
\end{table}

\begin{deluxetable}{lclc} 
\tablecolumns{4} 
\tablewidth{0pc} 
\tablecaption{Initial Abundances\label{abund}} 
\tablehead{ 
\colhead{Species}   & \colhead{Fractional Abundance} & \colhead{Species} & \colhead{Fractional Abundance}}
\startdata 
C &1.0(-08) &He      &6.0(-02) \\
O        &1.0(-07) &Fe        &1.0(-11) \\
Na        &1.0(-11) &S        &1.0(-08) \\
Si        &3.0(-13) &CO        &5.0(-05)\\ 
CS &3.0(-09) &CH$_{4}$       &3.0(-06)\\
H$_{2}$  &5.0(-01) &CN       &4.0(-09)\\
HCN       &3.0(-09) &HNC       &3.0(-09)\\
H$_{2}$O  &1.0(-08) &H$_{2}$S  &1.0(-09)\\
N$_{2}$        &1.0(-05) &NH$_{3}$       &1.0(-08)\\
O$_{2}$        &1.0(-08) &C$^+$      &1.0(-09)\\
He$^+$      &1.0(-11) &H$_{3}^+$   &6.0(-11)\\
S$^+$      &1.0(-11) &Fe$^+$   &1.0(-11)\\
Na$^+$   &1.0(-11) &HCO$^{+}$ &1.9(-09)\\
e$^{-}$ & 3.0(-09) &  (C)$_{\rm elem}$   &  5.3(-05)     \\
(O)$_{\rm elem}$ & 5.0(-05)  &       & \\
\enddata 
\tablecomments {a(-b) represents a$\times$10$^{-b}$.}
\end{deluxetable}

\begin{deluxetable}{lrrrrr}
\tablecolumns{6}
\tablewidth{0pc}
\tabletypesize{\scriptsize}
\tablecaption{Observed and Optimal Calculated Abundances in
L1527 \label{obs.v.calc}}
\tablehead{
\colhead{Species} & \colhead{Observed Abundance}   & \colhead{Reference}
 & \colhead{Calculated Abundance} &
\colhead{Without anions} & \colhead{Gas-Grain Model} }
\startdata
CO      &3.9(-05) &(1) &5.0(-5)  &5.0(-5) & 5.1(-05)\\
CN      &8.0(-11) &(1) &1.4(-9)  &1.1(-9) & 9.6(-10)\\
CS      &3.3(-10) &(1) &5.5(-10)  &5.9(-10) & 5.1(-09)\\
C$_{2}$S     &8.5(-11) &(2) &1.6(-10)  &1.7(-10) & 1.3(-10)\\
HCN     &1.2(-9) &(1) &1.3(-9)  &1.3(-9) & 7.7(-10)\\
HNC     &3.2(-10) &(1) &1.5(-9)  &1.4(-9) & 6.3(-10)\\
SO      &1.4(-10) &(1) &2.7(-10)  &3.3(-10) & 1.3(-10)\\
C$_{4}$H     &3.3(-9) &(2) &1.3(-10)  &5.7(-10) & 1.7(-09)\\
C$_{5}$H     &1.6(-11) &(2) &9.5(-11)  &1.7(-10) & 4.6(-10)\\
C$_{6}$H     &1.0(-11) &(2) &1.9(-11)  &4.8(-11) & 2.3(-10)\\
C$_{4}$H$^{-}$    &1.8(-13) &(3) &1.1(-11) &\nodata &\nodata\\
C$_{6}$H$^{-}$    &9.7(-13) &(4) &9.8(-12) &\nodata  &\nodata\\
HC$_{3}$N    &8.9(-10) &(1) &5.5(-10)  &5.8(-10) & 4.0(-10)\\
HC$_{5}$N    &9.7(-11) &(2) &1.1(-10)  &1.4(-10) & 1.6(-10)\\
HC$_{7}$N    &2.7(-11) &(2) &1.4(-11)  &1.8(-11) & 2.9(-11)\\
HC$_{9}$N    &2.5(-12) &(2) &2.1(-12)  &2.3(-12) & 5.9(-12)\\
H$_{2}$C$_{3}$    &1.0(-11) &(2) &7.9(-12)  &8.7(-12) & 1.1(-11)\\
H$_{2}$C$_{4}$& 2.2(-11) &(2,5) &1.7(-11)  &2.2(-11) & 1.2(-10)\\
H$_{2}$C$_{6}$     &2.5(-12) &(2,5) &3.0(-12)  &3.7(-12) & 2.5(-11)\\
C$_{3}$H$_{4}$    &7.8(-10) &(2) &1.4(-9)  &1.7(-9) & 2.2(-09)\\
HCO$^{+}$    &6.0(-10) &(1) &1.3(-9)  &1.2(-9) & 1.7(-09)\\
N$_{2}$H$^{+}$    &2.5(-10) &(1) &1.0(-11)  &1.0(-11) & 1.1(-11)\\
HCO$_{2}^{+}$   &1.0(-12) &(6) &1.9(-13)  &1.8(-13) & 4.1(-13)\\
\enddata
\tablenotetext{1}{\citet{jorgensen04}}
\tablenotetext{2}{\citet{sakai08a}}
\tablenotetext{3}{\citet{agundez08}}
\tablenotetext{4}{\citet{sakai08b}}
\tablenotetext{5}{Calculated carbene abundance was adjusted to 1\% of total C$_{n}$H$_{2}$ (see Section 3.4).}
\tablenotetext{6}{\citet{sakai08c}}
\tablecomments {a(-b) represents a$\times$10$^{-b}$.}
\end{deluxetable}

\begin{deluxetable}{lrrr} 
\tablecolumns{3} 
\tablewidth{0pc} 
\tablecaption{Predicted Optimal Fractional Abundances of Unobserved Gaseous Species   \label{predict}} 
\tablehead{ 
 \colhead{Species}    & \colhead{Gas-Phase Model} & \colhead{Gas-Grain Model}}
\startdata 
C$_{3}$H   &3.3(-10)   &1.0(-9)         \\
C$_{7}$H     &4.4(-12) &1.3(-10)\\
C$_{8}$H     &4.1(-12) &1.1(-10)\\
C$_{9}$H     &8.3(-13) &7.2(-11)\\
C$_{10}$H     &4.5(-13) & \nodata\\
C$_{5}$H$^{-}$    &3.0(-11) &\nodata\\
C$_{7}$H$^{-}$    &7.2(-12) &\nodata\\
C$_{8}$H$^{-}$    &1.9(-12) &\nodata\\
C$_{9}$H$^{-}$    &1.2(-12) &\nodata\\
C$_{10}$H$^{-}$    &2.0(-13) &\nodata\\
CH$_{4}$ & 2.9(-6) & 3.9(-6) \\
C$_{2}$H$_{2}$ &1.4(-8)  &3.7(-8)\\ 
H$_{2}$C$_{5}$ &8.0(-12)  &2.5(-11)\\
H$_{2}$C$_{7}$ &1.4(-12)  &7.1(-12)\\
H$_{2}$C$_{8}$ &5.8(-13) &5.6(-12)\\
H$_{2}$C$_{9}$ &2.6(-13) &2.6(-12)\\
CH$_{3}$C$_{4}$H &2.0(-10) &1.1(-9) \\
CH$_{3}$C$_{6}$H &1.4(-11) &1.7(-10)\\
HC$_{11}$N    &1.1(-13)  &\nodata\\
HC$_{13}$N    &5.2(-15) &\nodata \\
C$_{2}$N &2.2(-10) &4.3(-10)\\
C$_{3}$N &6.0(-11) &1.2(-10)\\
C$_{4}$N &4.3(-11) &1.7(-10)\\
C$_{5}$N &8.6(-11) &1.4(-10)\\
C$_{7}$N &1.0(-11) &2.6(-11)\\
C$_{9}$N &1.3(-12) &5.1(-12)\\
CH$_{3}$C$_{3}$N &1.5(-13)  &2.9(-13)\\
CH$_{3}$C$_{5}$N &7.5(-15) &2.3(-14)\\
CH$_{3}$C$_{7}$N &8.2(-16) &3.8(-15)\\
C$_{3}$N$^{-}$    &5.5(-12) &\nodata\\
N$_{2}$ &1.0(-5) &5.6(-6)\\
O$_{2}$ &7.4(-9) &1.1(-8)\\
H$_{2}$O & 7.4(-10) & 8.8(-10) \\
NH$_{3}$ & 7.0(-9) & 3.7(-9) \\
CO$_{2}$ &2.2(-9) &4.1(-9)\\
He$^{+}$ & 1.9(-12) & 8.8(-12) \\
C$^{+}$ & 2.9(-11) & 1.1(-10) \\
HCO$^{+}$ & 1.3(-9) & 1.7(-9) \\
H$_{3}^{+}$ & 2.7(-11) & 5.6(-11) \\
H$_{3}$O$^{+}$ & 8.8(-12) & 1.3(-11) \\
c-C$_{3}$H$_{2}$ & 4.2(-10) & 5.7(-10) \\
C$_{6}$H$_{6}$ &6.5(-11) &3.2(-10)\\
CH$_{2}$CN &4.4(-11) &7.4(-11)\\
H$_{2}$CO &5.4(-9) &4.3(-9)\\
HCOOH  &3.1(-12) &3.6(-12)\\
CH$_{2}$CO &1.2(-9) &1.0(-9)\\
CH$_{3}$CN &2.8(-12) &2.0(-12)\\
CH$_{3}$OH &3.4(-16) &5.0(-11)\\
HCOOCH$_{3}$ &1.8(-19) &3.8(-15)\\
CH$_{3}$OCH$_{3}$ &2.3(-17) &3.8(-16)\\ 
H$_{2}$CS &1.2(-9) &5.2(-10)\\
H$_{2}$S &7.7(-10) &3.3(-9)
\enddata 
\tablecomments {a(-b) represents a$\times$10$^{-b}$.}
\end{deluxetable}

\begin{deluxetable}{cccc} 
\tablecolumns{4} 
\tablewidth{0pc} 
\tablecaption{Dependence of Agreement with Observation on the Atomic Oxygen Abundance\label{ovary} } 
\tablehead{ 
\colhead{Initial fractional} & \colhead{$\kappa$}   & \colhead{Time of}    & \colhead{Number of Species}\\
\colhead{Oxygen Abundance} & \colhead{}   & \colhead{Best Agreement (yr)}    & \colhead{in Agreement} }
\startdata 
5.0(-08) &0.649&4.3(03) &18\\
1.0(-07) &0.652 &4.8(03) &18\\
2.0(-07) &0.652 &5.9(03) &19\\
5.0(-07) &0.636 &7.2(03) &20\\
1.0(-06) &0.621 &8.0(03) &18\\
\enddata 
\tablecomments {a(-b) represents a$\times$10$^{-b}$.}
\end{deluxetable}

\begin{deluxetable}{cccc} 
\tablecolumns{4} 
\tablewidth{0pc} 
\tablecaption{Dependence of Agreement with Observation on the Methane Abundance\label{methvary}} 
\tablehead{ 
\colhead{Initial fractional} & \colhead{$\kappa$}   & \colhead{Time of}    & \colhead{Number of Species}\\
\colhead{Methane Abundance} & \colhead{}   & \colhead{Best Agreement (yr)}    & \colhead{in Agreement} }
\startdata 
3.0(-07) &0.601 &8.8(03) &18\\
1.0(-06) &0.642 &6.5(03) &18\\
3.0(-06) &0.652 &4.8(03) &18\\
1.0(-05) &0.652 &4.3(03) &19\\
3.0(-05) &0.647 &4.3(03) &19\\
\enddata 
\tablecomments {a(-b) represents a$\times$10$^{-b}$.}
\end{deluxetable}

\clearpage

\begin{figure}
 \includegraphics[angle=0,width=.90\textwidth]{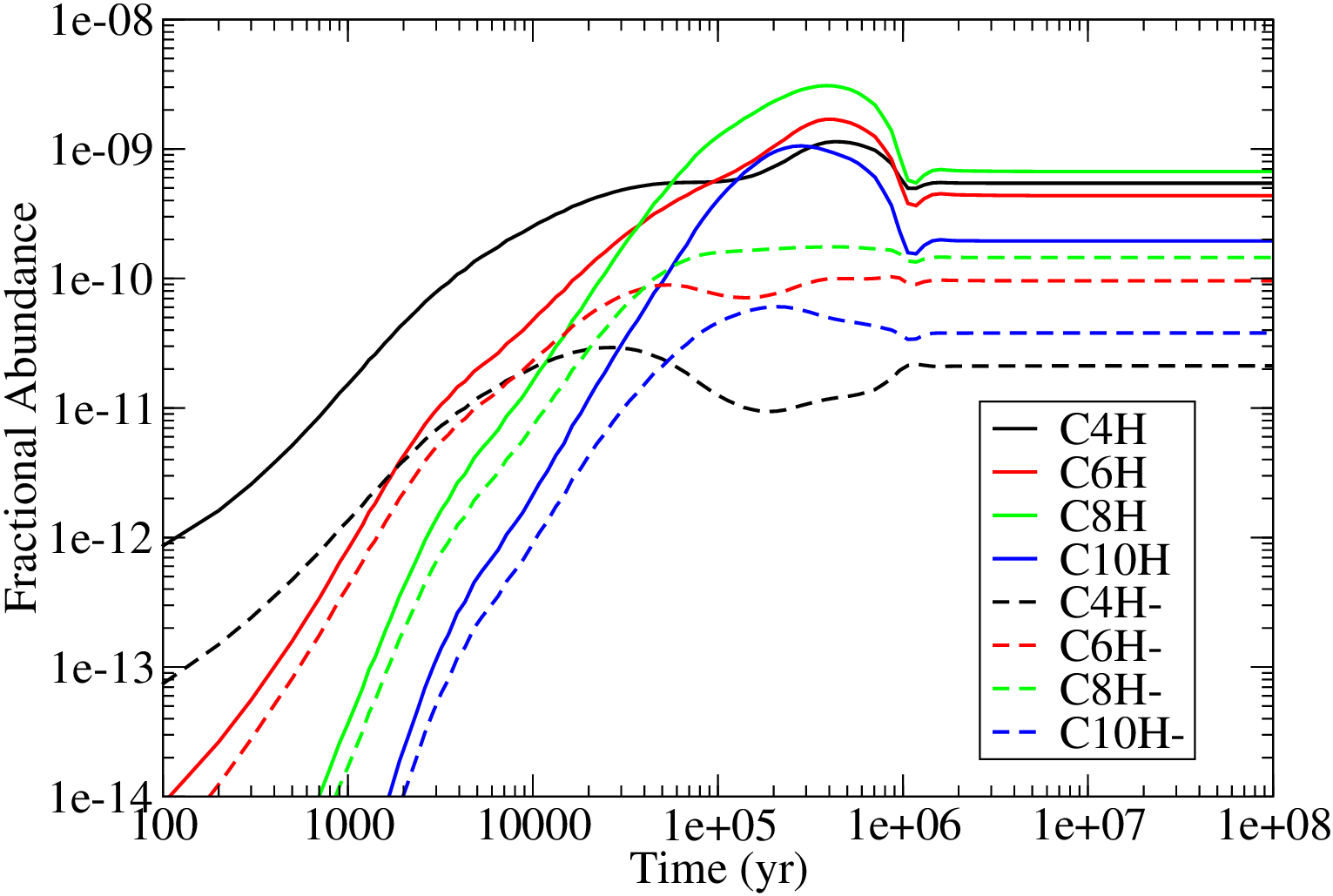}
\caption{The fractional abundances with respect to $n_{\rm H}$ for C$_{\rm 2n}$H and C$_{\rm 2n}$H$^{-}$ (n = 2-5) are plotted vs time for our model of L1527.\label{fig:cnh-}}
\end{figure}

\begin{figure}
 \includegraphics[angle=0,width=.90\textwidth]{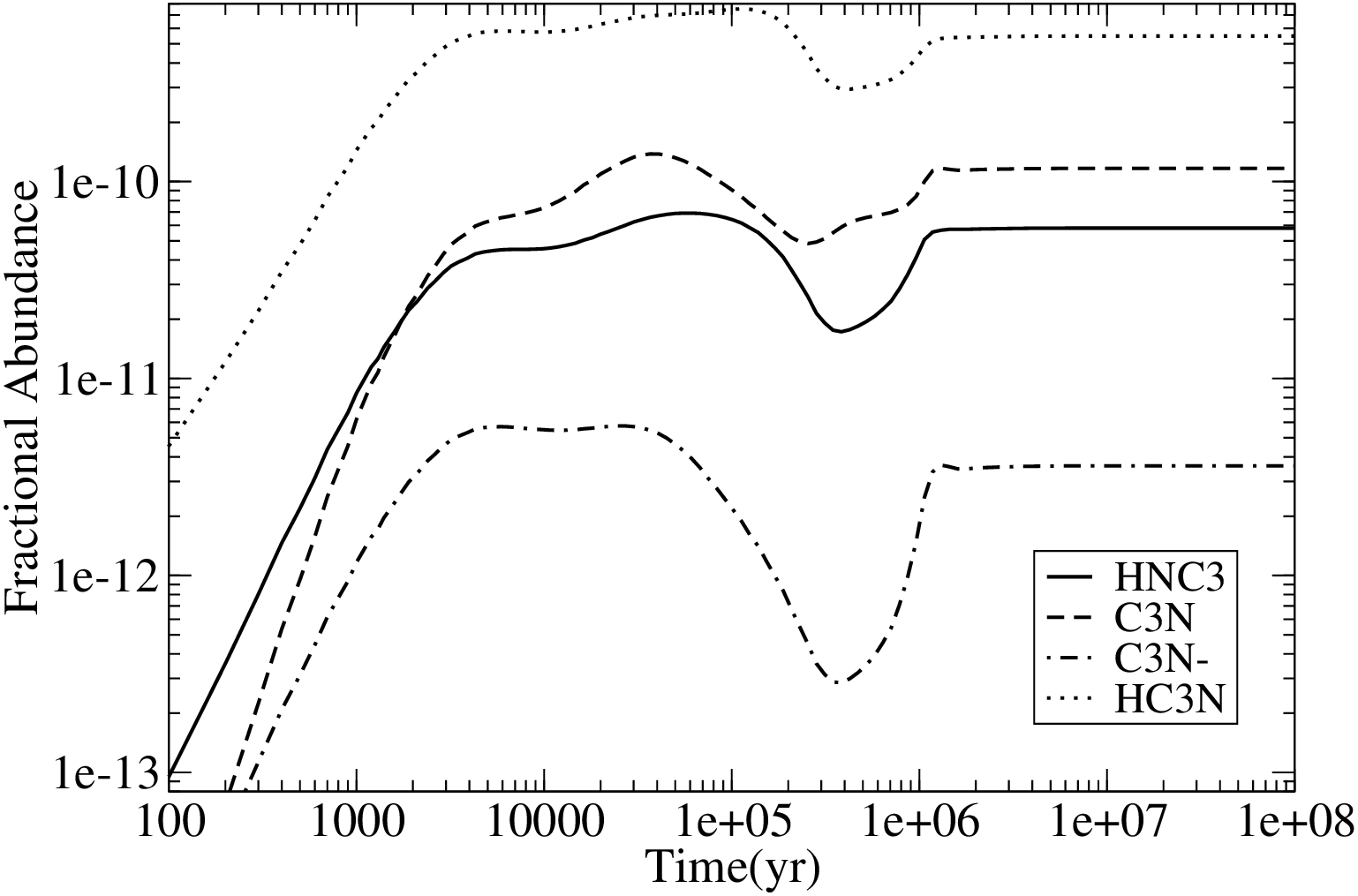}
\caption{The calculated fractional abundances with respect to $n_{\rm H}$ for C$_{3}$N$^{-}$, HCCCN, HNCCC, and C$_{3}$N are plotted against time for L1527.\label{fig:C3N-}}
\end{figure}

\begin{figure}
 \includegraphics[angle=0,width=.90\textwidth]{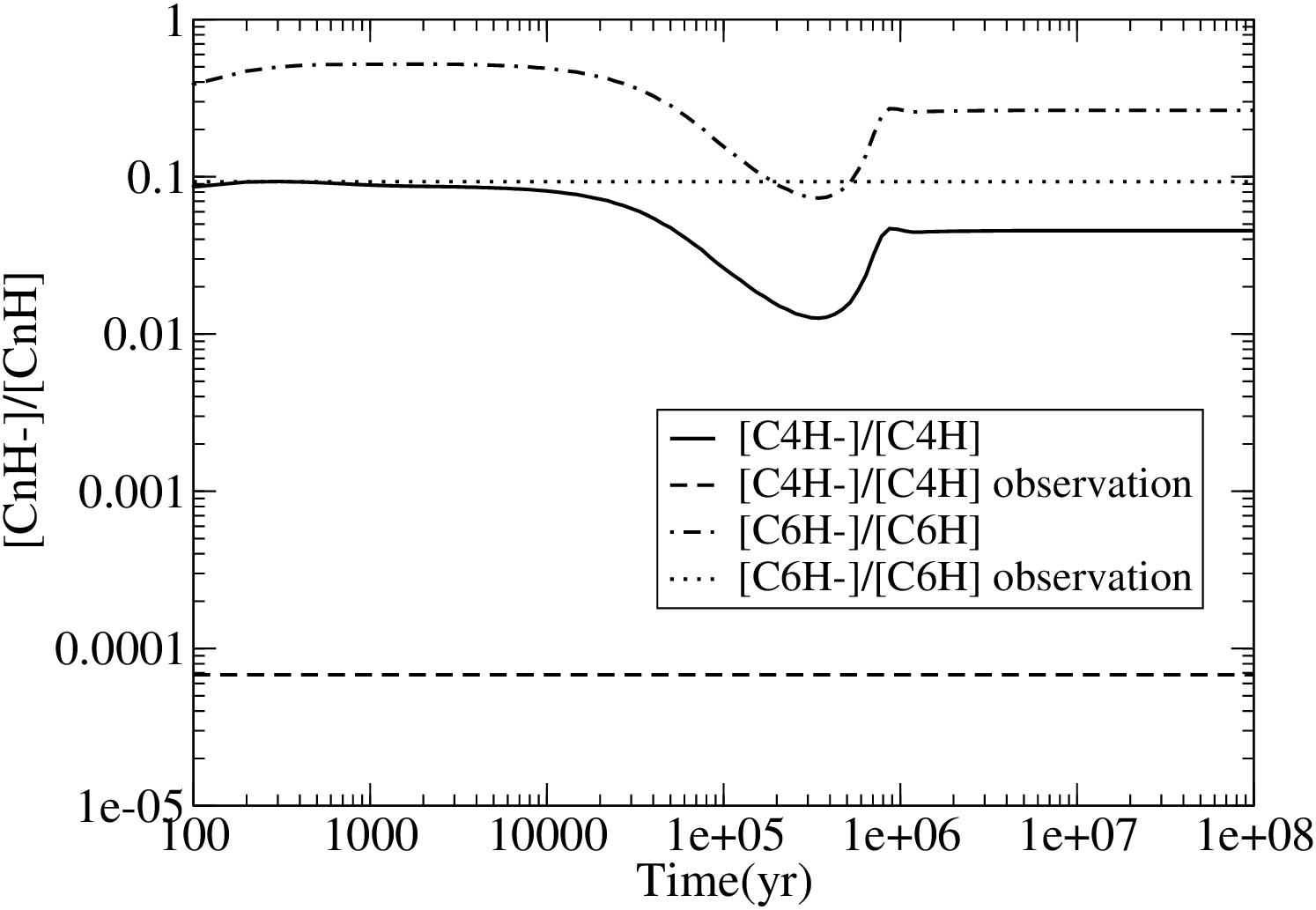}
\caption{The calculated anion-to-neutral abundance ratio for C$_{4}$H and C$_{6}$H as a function of time.  The observed values are plotted as horizontal lines.\label{fig:anion_ratio}}
\end{figure}

\begin{figure}
         \includegraphics[angle=0,width=.90\textwidth]{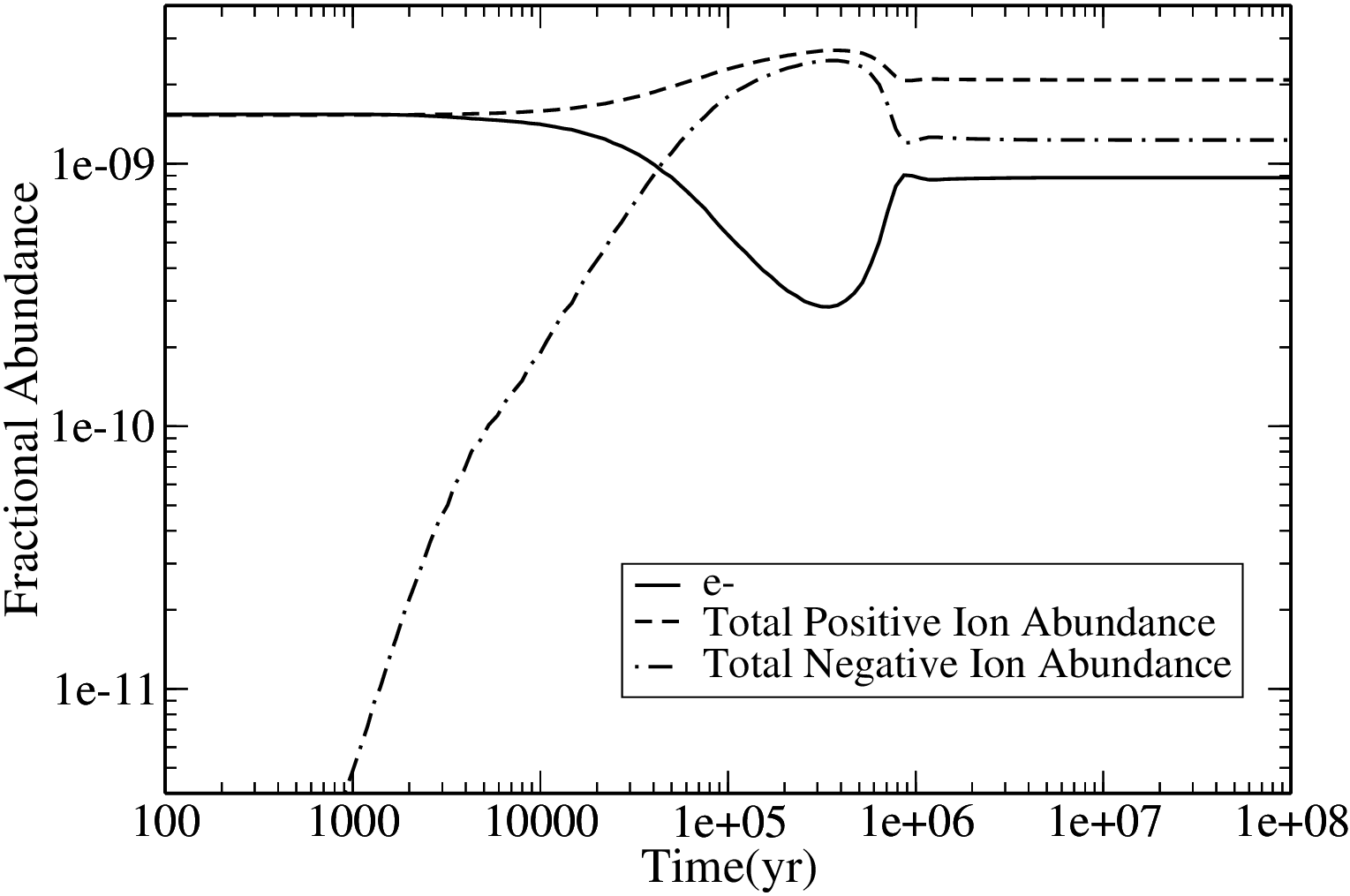}
\caption{The calculated fractional abundances of electrons, the total positive ions, and the total negative ions are plotted against time.  Note that the total negative ion abundance exceeds the electron abundance from a time of $3 \times 10^{4}$ yr. \label{fig:cmp_sum}}
\end{figure}

\clearpage
\begin{center}
         \includegraphics[angle=0,width=.90\textwidth]{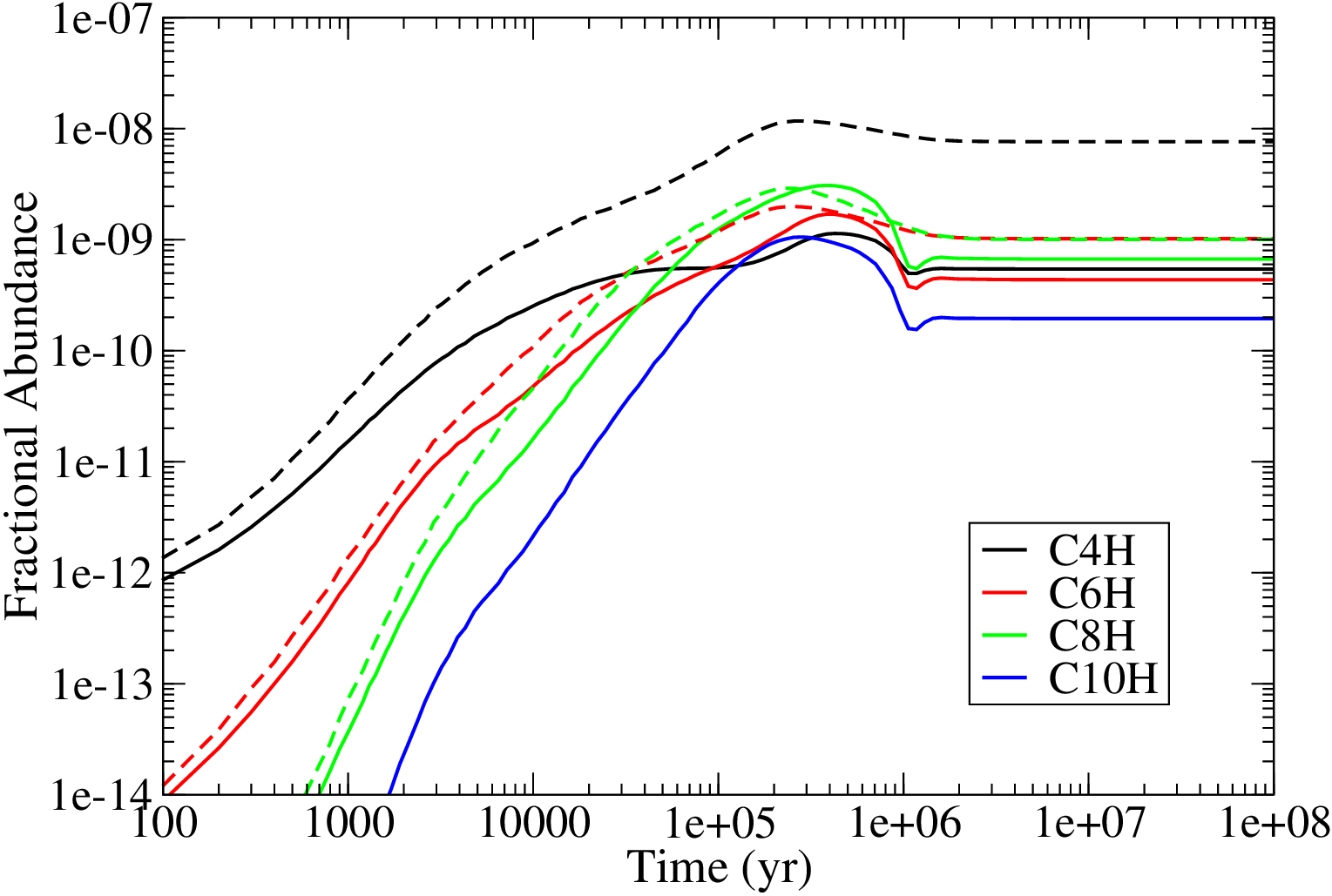}
\clearpage
         \includegraphics[angle=0,width=.90\textwidth]{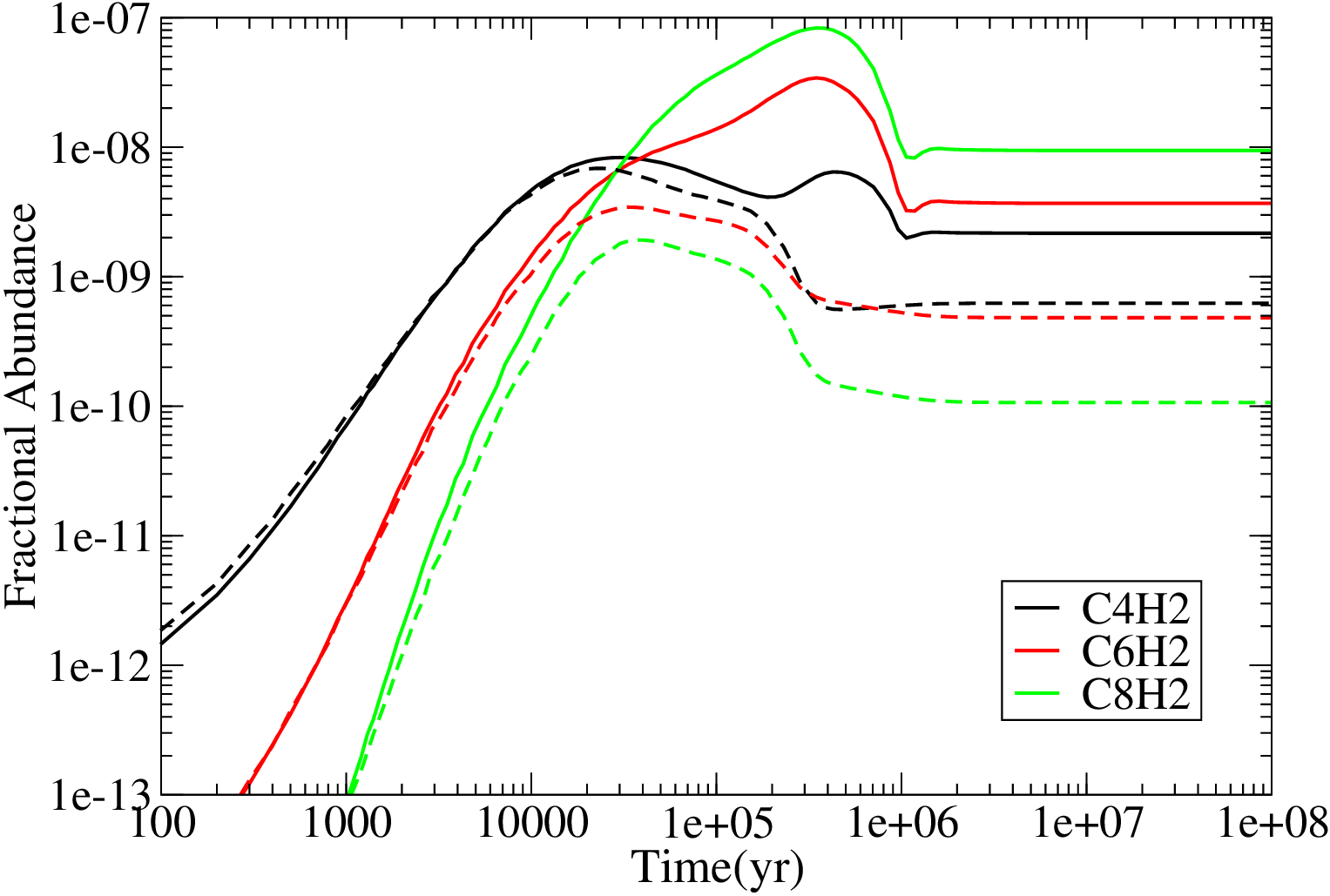}
\clearpage
          \includegraphics[angle=0,width=.90\textwidth]{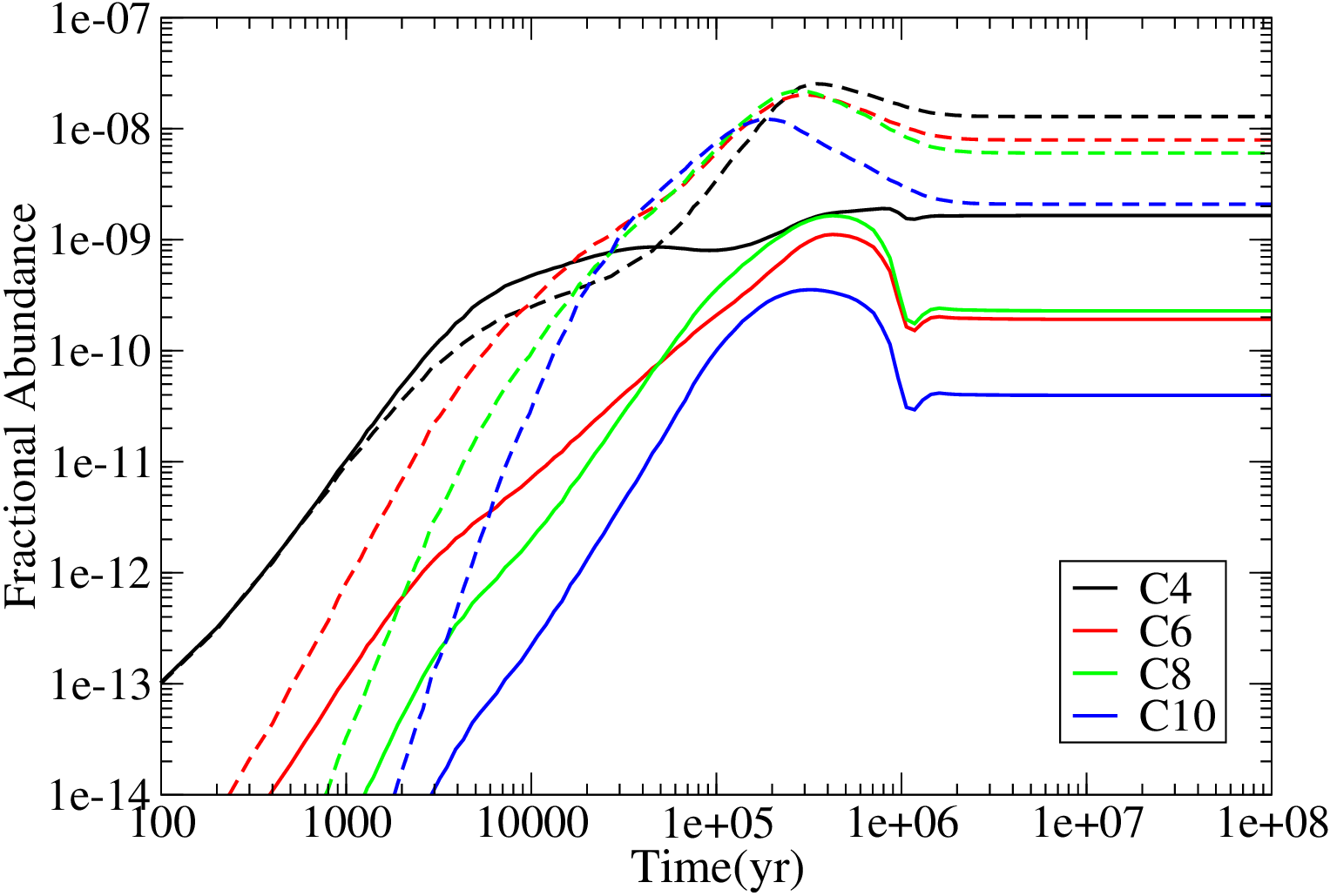}
\end{center}
\clearpage
\begin{figure}
\centering
          \includegraphics[angle=0,width=.90\textwidth]{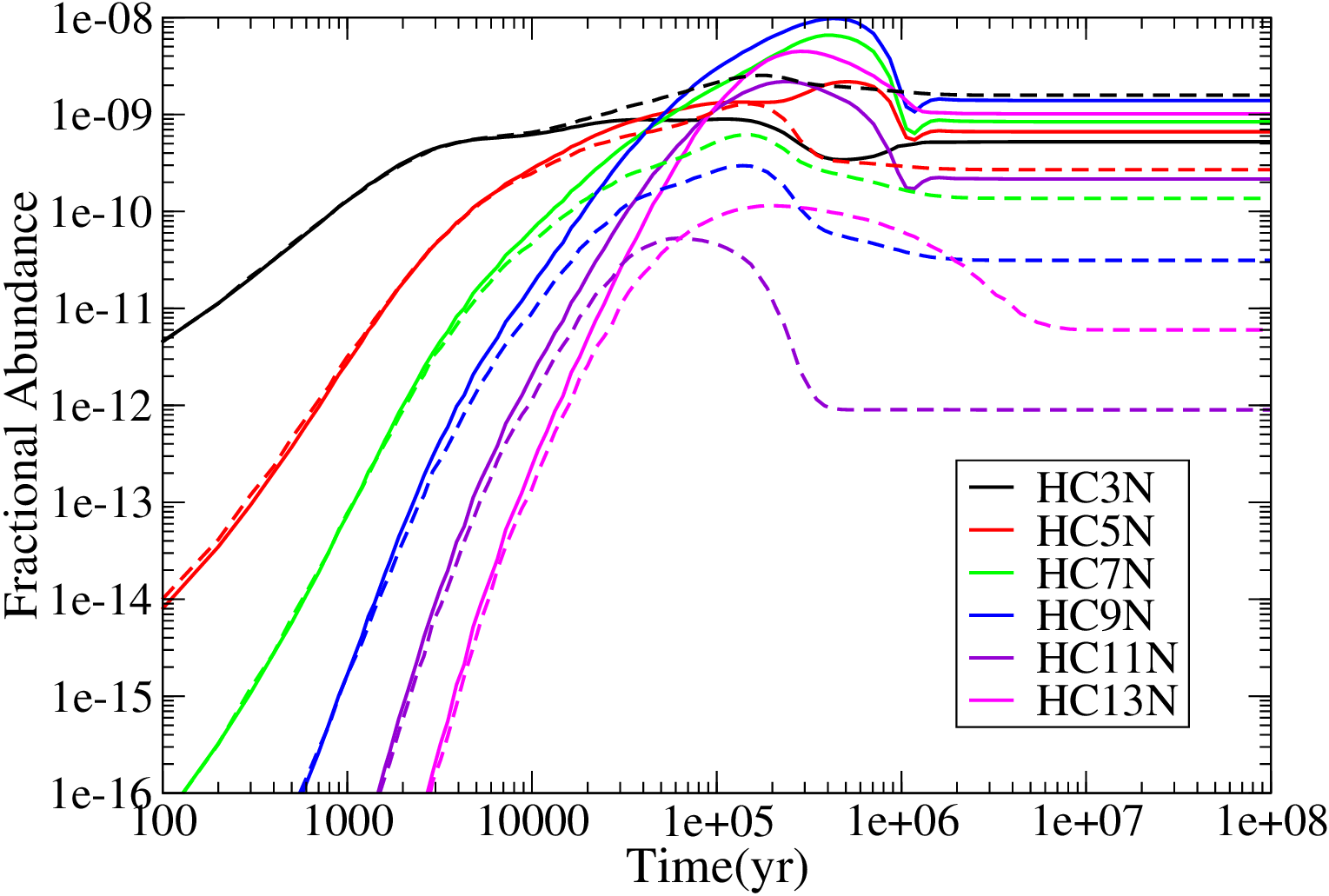}
\caption{Comparison of the time-dependent abundances of selected families of molecules --hydrocarbon radicals (C$_{\rm 2n}$H; Panel 1), hydrocarbons (C$_{\rm 2n}$H$_{2}$; Panel 2), carbon clusters (C$_{\rm 2n}$; Panel 3),  and cyanopolyynes (HC$_{\rm 2n+1}$N; Panel 4) -- with and without negative ions. Solid lines show the results with negative ions, while dotted lines show the results without negative ions.  Note that C$_{10}$H is not included in the network without anions. \label{fig:carbon_chain_l1527}}
\end{figure}

\end{document}